\documentstyle[ApJ,Times]{article}

\begin{document}

\def\simg{\mathrel{%
      \rlap{\raise 0.511ex \hbox{$>$}}{\lower 0.511ex \hbox{$\sim$}}}}
\def\siml{\mathrel{%
      \rlap{\raise 0.511ex \hbox{$<$}}{\lower 0.511ex \hbox{$\sim$}}}}
\def\Mesz{M\'esz\'aros~}
\def\ie{i.e$.$~} \def\eg{e.g$.$~} \def\etal{et al$.$~} \def\dd{{\rm d}}
\def\adindex{\hat{\gamma}} \def\eq{eq$.$~} \def\eqs{eqs$.$~} \def\deg{^{\rm o}}

\title{Dynamical Evolution, Light-Curves and Spectra of Spherical and Collimated 
       Gamma-Ray Burst Remnants}

\author{A. Panaitescu \& P. \Mesz}
\affil{Department of Astronomy \& Astrophysics,
    Pennsylvania State University, University Park, PA 16802}

\begin{abstract}

 We present an analytical approach to the dynamical evolution of fireballs or 
axially symmetric jets expanding into an external medium, with application to 
Gamma-Ray Burst remnants. This method leads to numerical calculations of fireball 
dynamics that are computationally faster than hydrodynamic simulations. 
It is also a very flexible approach, that can be easily extended to include more 
complex situations, such as a continuous injection of energy at the reverse shock, 
and the sideways expansion in non-spherical ejecta. 
 
 Some features of the numerical results for the remnant dynamical evolution are 
discussed and compared to the analytical results. We find that the ratio of the
observer times when the jet half-angle reaches twice its initial value and when
the light cone becomes wider than the jet is substantially smaller than predicted 
analytically. The effects arising from the remnant curvature and thickness further 
reduce this ratio, such that the afterglow light-curve break due to the remnant 
finite angular extent overlaps the weaker break due to the jet's sideways expansion. 

 An analysis of the effects of the curvature and thickness of the remnant on the 
afterglow light-curves shows that these effects are important and should be taken 
into account for accurate calculations of the afterglow emission. 

\end{abstract}


\section{Introduction}

 Previous work on the dynamics of Gamma-Ray Burst (GRB) remnants was done analytically 
(Sari 1997, Vietri 1997, Wei \& Lu 1998, Chiang \& Dermer 1999, Rhoads 1999) or was based 
on hydrodynamical codes (Panaitescu \& \Mesz 1998, Kobayashi, Piran \& Sari 1999). 
Analytical treatments of the major afterglow features (\Mesz \& Rees 1997, Vietri 1997, 
Waxman 1997, Wijers, Rees \& \Mesz 1997, \Mesz, Rees \& Wijers 1998, Sari, Piran \& 
Narayan 1998, Rhoads 1999) have used asymptotic solutions where the remnant Lorentz 
factor is a power-law in radius and the post-shock energy density is determined only 
by the ejecta bulk Lorentz factor. The advantages and drawbacks of the analytical and 
numerical approaches are evident: the former may not lead to sufficiently accurate results, 
but is more flexible and sometimes more powerful, while the second provides accuracy at 
the expense of substantial computational efforts, and may lack flexibility in including 
more sophisticated features. 

 In this work we further develop the semi-analytical approach presented by 
Panaitescu, \Mesz \& Rees (1998) (PMR98), with the aim of obtaining a method 
that allows fast computations of remnant dynamics, without making recourse to 
laborious hydrodynamic numerical codes, while retaining enough physical details 
to allow an accurate calculation of the afterglow light-curves and spectra.

\section{Remnant Dynamics}

\subsection{Model Assumptions and Features} 

 We assume that, at any time during its evolution, the remnant is axially symmetric, 
\ie there are no angular gradients in the shocked fluid, and all relevant physical 
parameters are functions only of the radial coordinate. This is equivalent to assuming 
that the initial energy distribution in the ejecta and the external gas are isotropic 
and that, in the case of a jet-like ejecta, the physical parameters of the remnant 
respond on a short time-scale to the effect of sideways expansion of the shocked gas. 
We also assume that the Lorentz factor within the shocked fluid is constant. Other  
features taking into account in the model are:
$i)$ a delayed energy injection (Rees \& \Mesz 1998), where for definiteness we shall 
  use a power-law energy injection, 
$ii)$ an inhomogeneous external fluid (Vietri 1997, \Mesz \etal 1998); for simplicity 
  we consider an external density which is a power-law in the radial coordinate, 
  which includes the homogeneous and the constant velocity pre-ejected wind cases, and 
$iii)$ the above mentioned sideways expansion, if the ejecta are not spherical (Rhoads 1999). 
Each of these features brings or modifies a specific term in the differential equations 
shown below, which give the evolution of the mass, kinetic and internal energy of the 
remnant.

\subsection{External Medium} 

 In the next sections we shall consider only homogeneous external media, nevertheless 
the model we developed is applicable for a more general case, where the external gas 
density varies as a power-law with the radius: $\rho_{ex}(r) = \rho_d (r/r_d)^{-\alpha}$, 
$r_d$ being the deceleration radius, defined as the radius at which the fastest (initial) 
part of the ejecta, moving with Lorentz factor $\Gamma_0$, sweeps up an amount of external 
gas equal to a fraction $\Gamma_0^{-1}$ of its own mass $M_0=E_0/\Gamma_0 c^2$:
\begin{equation}
 r_d = \left[\frac{(3-\alpha) E_0}{\Omega_0 n_d m_p c^2 \Gamma_0^2}\right]^{1/3} \;,
\label{rdec}
\end{equation}
$E_0$ being the energy of the leading ejecta, $\Omega_0$ the solid angle of the cone within 
which it was released and $n_d$ the external particle density at $r_d$. The $r_d$ is the 
radius at which the deceleration of the fireball due the interaction with the external gas 
becomes important. Note that $r_d$ has only a weak dependence on $\alpha$.

The continuous interaction with the external gas increases the remnant mass 
\begin{equation}
 [\dd M]_{ex} = \Omega(r) \rho_{ex}(r) r^2 \dd r \;,
\label{dMex}
\end{equation}
where $\Omega(r)$ is the remnant solid angle at radius $r$, which is larger than the initial 
$\Omega_0$ due to the sideways expansion. The evolution of the half-angle $\theta$ of the 
remnant is $\dd \theta = c_s \dd t'/r$, where $c_s=c/\sqrt{3}$ is the comoving frame sound 
speed and $\dd t'= (\beta c \Gamma)^{-1} \dd r$ is the comoving time, $\Gamma$ and $\beta$
being the Lorentz factor and speed of the shocked ejecta. Using $\Omega = 2\pi (1-\cos \theta)$, 
the remnant solid angle evolves as
\begin{equation}
r \frac{\dd \Omega}{\dd r} =\sqrt{\frac{\Omega(4\pi-\Omega)}{3(\Gamma^2-1)}}\;.
\label{dOmega}
\end{equation}
The interaction of the remnant with the external medium decelerates the remnant 
and heats it. Using energy and momentum conservation for the interaction 
between the remnant and the infinitesimal swept-up mass $[\dd M]_{ex}$, one 
obtains that the comoving frame internal energy of the newly shocked gas is 
($\Gamma-1$) times larger than its rest-mass energy, and the changes in the 
remnant internal energy $U$ and total kinetic energy $K$ are given by
\begin{equation}
 [\dd U]_{ex} = A (\Gamma-1)c^2 [\dd M]_{ex} \;,
\label{dUex}
\end{equation}
\begin{equation}
 [\dd K]_{ex} \equiv (Mc^2+U)[\dd \Gamma]_{ex} = -(\Gamma^2-1)c^2 [\dd M]_{ex} \;.
\label{dGex}
\end{equation} 
The multiplying factor $A$ in the right-hand side of equation (\ref{dUex}) was 
introduced to account for possible radiative losses in the shocked external gas. 
It represents the fraction of the internal energy of the shocked external gas 
that is not radiated away. Thus $A=0$ corresponds to a fully radiative remnant, 
which implies radiative electrons and strong coupling between electrons and 
protons, while $A=1$ is for a fully adiabatic remnant.

\subsection{Delayed Energy Injection}

 As suggested by Rees \& \Mesz (1998) it is possible that the initial instantaneous 
deposition of energy in the fireball is not uniform in the entire ejecta, in the 
sense that some parts of it have been given more energy and have been accelerated 
to higher Lorentz factors, the slower ejecta
catching up with the faster ones as these are decelerated by the interaction 
with the surrounding medium. The entire process is fully characterized by the 
energy distribution $(\dd E/\dd \Gamma)_{inj}$ in the ejecta at the end of the 
initial phase of acceleration, with all other relevant quantities resulting 
from the kinematics and energetics of the ``catching up". The velocity $\beta_f$
of the incoming ejecta satisfies $r=\beta_f ct$ (free expansion), where $t$ is
the lab-frame time: $t = t_d + \int_{r_d}^r [c\beta(r)]^{-1} \dd r$,
with $\beta(r)$ the speed of the decelerated ejecta. We assume that there is a
significant delayed energy injection only at $r > r_d$, so that equation (\ref{rdec})
remains valid. Thus the Lorentz factor $\Gamma_f$ 
of the delayed ejecta entering the decelerated part of the fireball is given by
\begin{equation}
   \frac{\dd \Gamma_f}{\dd r} = - \left(\frac{\beta_f}{\beta}-1\right)
                                \frac{\Gamma_f^3 \beta_f}{ct} \;.
\label{Gf}
\end{equation}
Using the energy and momentum conservation, one can calculate the increase in the 
remnant kinetic and internal energies as result of the delayed injection:
\begin{equation}
 [\dd U]_{inj} = [\Gamma\Gamma_f (1-\beta\beta_f) - 1]\,c^2 [\dd M]_{inj} \;,
\label{dUinj}
\end{equation}
\begin{equation}
 [\dd K]_{inj} \equiv (Mc^2+U)[\dd \Gamma]_{inj} = 
     \Gamma_f[1 -\Gamma^2(1-\beta\beta_f)]\,c^2 [\dd M]_{inj} \;,
\label{dGinj}
\end{equation}
where
\begin{equation}
  [\dd M]_{inj} = \left(\frac{\dd E}{\dd \Gamma}\right)_{inj} 
 \frac{|\dd \Gamma_f|}{\Gamma_f-1} = F(\Gamma_f,\Gamma) M_{INJ} \frac{\dd r}{r}\;,
\label{dMinj}  
\end{equation}
is the infinitesimal injected mass, $F(\Gamma_f,\Gamma)$ being a function that 
depends on the details of the delayed energy injection, and $M_{INJ}$ the total 
mass of the delayed ejecta (see PMR98). In the numerical calculations we shall 
consider the particular case of a power-law injection: $(\dd E/\dd \Gamma)_{inj} 
\propto \Gamma_f^{-s}$ (Rees \& \Mesz 1998) for $\Gamma_m < \Gamma_f < \Gamma_d$, 
where $\Gamma_m$ and $\Gamma_d \equiv \Gamma (r=r_d)$ are the minimum and maximum 
Lorentz factors of the delayed ejecta.

\subsection{Adiabatic Cooling. Remnant Volume} 

 The delayed energy input at the reverse shock that moves into the incoming ejecta 
and the heating of the external fluid by the forward shock increase the internal 
energy of the remnant. This energy is lost adiabatically and radiatively. If they 
were acting alone, adiabatic losses would accelerate the remnant; in the presence 
of the external fluid, they re-convert internal energy into kinetic energy, 
which mitigates the remnant deceleration. This is described quantitatively by
\begin{equation}
 [\dd U]_{ad} = -(\adindex-1) (\dd_r V'/V') U \;,
\label{dUad}
\end{equation}
\begin{equation}
 [\dd K]_{ad} \equiv (Mc^2+U) [\dd \Gamma]_{ad} = -\Gamma  [\dd U]_{ad} \;,
\label{dGad}
\end{equation}
where $V'=V'_{RS}+V'_{FS}$ is the comoving volume of the shocked ejecta (located 
behind the reverse shock) and of the swept-up external gas (behind the forward 
shock), and $\adindex$ is the adiabatic index of the remnant gas ($\adindex=4/3$ 
for hot gas). In equation (\ref{dUad}) $\dd_r V'$ denotes the infinitesimal 
variation of the comoving volume due only to the radial expansion of the gas, 
excluding the infinitesimal increases due to the addition of shocked fluid and 
to the sideways expansion (we neglect the adiabatic losses due to the sideways 
expansion, representing the acceleration of the outer parts of the fluid in the 
direction perpendicular to the radial direction of the flow). 
 
 At this point we need a prescription for calculating $\dd_r \ln V'$. For the 
the radiative losses one also has to calculate the comoving volume $V'$, to 
determine the comoving energy density, necessary for the computation of the 
magnetic field.  We consider two models: {\sl Model 1}, where we assume that 
the lab frame increase in the thickness of the shocked fluid (external or delayed 
ejecta) is due only to the addition of new gas, and {\sl Model 2}, where we assume 
that the comoving density of the two shocked fluids are uniform behind each shock, 
and have the values set by the shock jump equations. Only {\sl Model 1} is 
consistent with the assumption that $\Gamma$ is constant in the entire shocked 
fluid, as this assumption implies that the remnant is neither dilating nor 
contracting in the radial direction.  {\sl Model 2} implies the existence of
a velocity gradient in the shocked fluid, nevertheless this gradient
is expected to be small and the assumption of constant $\Gamma$ within the
remnant may still be safely used for the calculation of the relativistic
effects.

\subsubsection{Model 1} 

 The comoving volume is $V' = \Omega r^2 \Gamma \Delta$, where $\Delta$ is the       
lab-frame thickness of the remnant, determined by the relative motion of the forward 
and reverse shocks: $\dd \Delta = (\beta_{FS} - \beta_{RS}) \dd t$. The lab-frame 
reverse shock speed $\beta_{RS}$ can be calculated with a Lorentz transformation from 
$\beta'_{RS}$, the speed of the reverse shock measured in the frame of the ejecta
entering this shock.  The assumption that the frame thickness of the already shocked 
fluid remains constant (or equivalently that the flow velocity is uniform within the 
remnant) implies that $\dd_r \ln V' = \dd \ln (\Gamma r^2)$, 
therefore
\begin{equation}
 \frac{\dd_r V'}{V'} \stackrel{M1}{=} 2 \frac{\dd r}{r} + 
           \frac{\dd \Gamma}{\Gamma} \;,
\label{dV}
\end{equation}
and allows the calculation of $\beta'_{RS}$ and $\beta_{FS}$ from $\Gamma$, the
remnant bulk Lorentz factor: 
\begin{equation}
 \beta'_{RS} = \frac{(\Gamma'-1)(\adindex\Gamma'+1)}
               {\beta'\Gamma'[\adindex(\Gamma'-1)+1]} \;, 
 \beta_{FS} = \frac{(\Gamma-1)(\adindex\Gamma+1)}
               {\beta\Gamma[\adindex(\Gamma-1)+1]} \;. 
\end{equation}    
where $\Gamma'= \Gamma \Gamma_f (1-\beta \beta_f)$ is the Lorentz factor of
the shocked delayed ejecta measured in the frame of the incoming ejecta located
just ahead of the reverse shock. After all the delayed ejecta has caught up with 
the remnant, or if there is no such delayed injection, $\beta_{RS}$ is set equal 
to $\beta$, the speed of the contact discontinuity.

\subsubsection{Model 2} 

 In this model, the two volumes $V'_{RS}$ and  $V'_{FS}$ can be calculated from 
the masses of the shocked gases (\eqs [\ref{dMex}] and [\ref{dMinj}]) and from 
the comoving densities, assumed to be uniform in each shell and having the same 
value as in the proximity of the shocks. The comoving densities $\rho'_{RS}$ and 
$\rho'_{FS}$ behind the reverse and forward shock are determined 
by the comoving densities of the un-shocked fluids, $\rho'_f$ and $\rho_{ex}$, 
and by the Lorentz factors of the shocked gases, $\Gamma'$ and $\Gamma$:
\begin{equation}
 \rho'_{RS} = \frac{\adindex \Gamma'+1}{\adindex -1} \rho'_f\;, \quad
 \rho'_{FS} = \frac{\adindex \Gamma +1}{\adindex -1} \rho_{ex} \;. 
\label{rho}
\end{equation} 

{\bf Reverse Shock.} 
The comoving density of the ejecta ahead of the reverse shock can be calculated by 
equating $[\dd M]_{inj}$ given by equation (\ref{dMinj}) with the mass $\Gamma_f 
\rho'_f \Omega_f r^2 \dd l$ swept up by the reverse shock as the remnant moves 
from $r$ to $r+\dd r$, where $\dd l = ct |\dd \beta_f| = (\beta_f/\beta -1) \dd r$ 
is the infinitesimal lab frame distance relative to the contact discontinuity covered 
by the reverse shock, and $\Omega_f (\Gamma_f,r) \siml \Omega$ is the solid angle of 
the delayed ejecta. The assumption of isotropy within the ejecta beaming 
cone does not hold in the delayed energy input case, our 1-dimensional model 
including the delayed energy injection being applicable only to spherical ejecta for 
any lab-frame times and to beamed ejecta for times when the sideways expansion is 
negligible: $\Omega_f \sim \Omega \sim \Omega_0$. The end result is 
$\rho'_f = (\Gamma_f^2 \beta_f / \Omega_0 r^2 ct) (\dd M/\dd \Gamma)_{inj}$, so 
$V'_{RS} = (M_0+ \int_{r_d}^r [\dd M]_{inj})/ \rho'_{RS}$ can be calculated by 
integrating equation (\ref{dMinj}) and using the first equation (\ref{rho}). 
From $\dd_r (\ln V'_{RS}) = -\dd (\ln \rho'_{RS})$ and equations (\ref{Gf}) and 
(\ref{rho}), one can write
\begin{equation}
 \frac{\dd_r V'_{RS}}{V'_{RS}} \stackrel{M2}{=} G(\Gamma_f,\Gamma)
    \frac{\dd r}{r} + \frac{\adindex \Gamma_f}{\adindex \Gamma'+1}
         \left(\frac{\beta_f}{\beta}-1\right) \dd \Gamma \;,
\end{equation}
where $G(\Gamma_f,\Gamma)$ is a function of the details of the delayed injection.
After the end of the delayed injection the comoving volume of the shocked delayed
ejecta is considered constant.

{\bf Forward Shock.} 
The volume of the shocked external fluid $V'_{FS} = (M_0 \Gamma_0^{-1} + \int_{r_d}^r 
[\dd M]_{ex})/\rho'_{FS}$ can be calculated using the second equation (\ref{rho}) 
and the swept-up mass obtained by integrating equation (\ref{dMex}). From the second 
equation (\ref{rho}) it can be shown that
\begin{equation}
  \frac{\dd_r V'_{FS}}{V'_{FS}} \stackrel{M2}{=} \alpha \frac{\dd r}{r} - 
                \frac{\adindex} {\adindex \Gamma +1} \dd \Gamma \;.
\label{dVFS}
\end{equation}

\subsection{Differential Equations for Remnant Dynamics}

 The remnant dynamics is given by the differential equations describing the 
evolution of the total kinetic and internal energies, coupled through the
adiabatic losses:
\begin{equation}
 \dd U =  [\dd U]_{inj}+[\dd U]_{ad}+[\dd U]_{ex} \;,
\label{dU}
\end{equation}
\begin{equation}
 \dd K \equiv (Mc^2+U) \dd \Gamma = [\dd K]_{inj}+[\dd K]_{ad}+[\dd K]_{ex}\;,
\label{dG}
\end{equation}
where the quantities in the right-hand side terms are given by equations (\ref{dUex}), 
(\ref{dGex}), (\ref{dUinj}), (\ref{dGinj}), (\ref{dUad}) and (\ref{dGad}). 
By substituting the term $[\dd U]_{ad}$ from equation (\ref{dG}) in equation (\ref{dU}) 
($[\dd U]_{ad}$ appears in the expression of $[\dd K]_{ad}$ -- see \eq [\ref{dGad}]), 
one arrives at $\dd [M(\Gamma-1)+ \Gamma U] = (\Gamma_f-1) [\dd M]_{inj}$, which simply 
states that the net variation of the total energy of the adiabatic remnant equals the 
input of energy through the delayed injection (global energy conservation). 

 With the aid of all the relevant equations previously derived, equations (\ref{dU}) and 
(\ref{dG}) can be used to calculate $\dd \Gamma/\dd r$ and $\dd U/\dd r$, \ie the evolution 
of the flow Lorentz factor and of the co-moving internal energy. These equations are solved 
numerically together with equation (\ref{Gf}) for $\Gamma_f$, equation (\ref{dOmega}) 
for $\Omega$, and the differential equation for the remnant mass resulting from 
equations (\ref{dMex}) and (\ref{dMinj}):
\begin{equation}
 r \frac{\dd M}{\dd r} = 
     F(\Gamma_f,\Gamma) M_{INJ} + (3-\alpha) \frac{\Omega}{\Omega_0} 
     \left(\frac{r}{r_d}\right)^{3-\alpha} \frac{M_0}{\Gamma_0} \;.
\label{dM}
\end{equation}
The initial conditions for the set of differential equations for remnant 
dynamics are given by the values of the relevant quantities at $r=r_d$ (see 
PMR98): $\Gamma_f=\Gamma=0.62 \Gamma_0, U=0.62\, A M_0c^2, M=(1+\Gamma_0^{-1}) 
M_0, \Omega \sim \Omega_0$ .

\section{Numerical Results for the Remnant Dynamics}

 The remnant Lorentz factor $\Gamma$ and the internal energy $U$ determine
the electron random Lorentz factor and the magnetic field, both necessary for 
the calculation of the afterglow emission. Thus we are interested in solving the 
remnant differential equations to calculate the evolution of $\Gamma$ and $U$ with 
the observer time $T$, 
\begin{equation}
 \dd T = (1+z) (1-\beta) \dd t = 
      (1+z) \left( \frac{\Gamma}{\sqrt{\Gamma^2-1}} -1 \right) \frac{\dd r}{c} \;,
\label{Time}
\end{equation}
where $z$ is the source redshift.  Equation (\ref{Time}) gives the time $T_{CD}$ 
when the radiation emitted along the line of 
sight toward the observer and from the contact discontinuity arrives at Earth. 
If most of the radiation comes from the fluid close to forward shock then it is 
necessary to calculate the observer time using the Lorentz factor of this shock.
For a relativistic remnant $\Gamma_{FS} \sim \sqrt{2} \Gamma$, thus $T_{FS} = 
T_{CD}/2$. 

 Equations (\ref{rdec}) and (\ref{dG}) show that the remnant dynamics is determined 
by $\varepsilon_0 \equiv E_0/\Omega_0$, the energy per solid angle in the ejecta, 
the jet initial solid angle, which determines when the jet sideways expansion becomes
important, the parameters $n_d$ and $\alpha$ characterizing the surrounding 
medium, the remnant initial Lorentz factor $\Gamma_0$ and the remnant radiative
efficiency. In the case of an adiabatic remnant running into a homogeneous external 
medium, $\Gamma_0$ cancels out from the expression for $\Gamma (T)$, thus it is an 
irrelevant parameter, from the observer's point of view. However $\Gamma_0$ is 
an important parameter for a radiative remnant, or if the external medium is not 
homogeneous. The remnant dynamics is also determined by the parameters of the 
delayed energy injection, which for a power-law injection are $\Gamma_m$, $\Gamma_0$, 
$s$ and $M_{INJ}$ (or, equivalently, the entire injected energy $E_{INJ}$).
The effect of some of these parameters can be assessed from Figures 1--3.

\subsection{Spherical Remnants}

 Figure 1 shows the evolution of $-\dd \log \Gamma/\dd \log r$ for a spherical 
remnant with no delayed energy input, running into a homogeneous external medium. 
The non-relativistic phase, defined by $\Gamma < 2$, sets in at $r < 10\, r_d$ for a 
fully radiative remnant and at $r < 100\, r_d$ for an adiabatic one. An analytical 
treatment of the remnant dynamics leads to $-\dd \log \Gamma/\dd \log r=(3-\alpha)/(1+A)$, 
as long as the remnant is relativistic. Thus, if $\alpha=0$, $\Gamma \propto r^{-3}$ 
for a radiative remnant and $\Gamma \propto r^{-3/2}$ for an adiabatic one. These 
results hold for $r_d \ll r \ll \Gamma_0^{1/3} r_d$ in the former case and for 
$r_d \ll r \ll \Gamma_0^{2/3} r_d$ in the latter. The values shown in Figure 1 at
early times (\ie in the relativistic phase), are consistent with the analytical 
expectations. Due to the fact that the $r^{-3}$ phase is short lived for a radiative
remnant, this regime is not strictly reached for the case shown in Figure 1 
($\Gamma_0 = 500$), where the steepest $\Gamma$-decay attained is $\propto r^{-2.85}$. 
Only Lorentz factors $\Gamma_0 > 10^3$ allow this phase to fully develop at very 
early observer times ($T < 0.1$ day).  In the case of a pre-ejected wind ($\alpha=2$), 
we obtained numerically the analytical results $\Gamma \propto r^{-1/2}$ and $\Gamma
\propto r^{-1}$ for an adiabatic and a radiative remnant, respectively (these cases are 
not shown in Figure 1).

 Figure 2 shows the effect of a delayed energy input on the dynamics of an adiabatic 
remnant, assuming a homogeneous external gas and a power-law distribution of energy 
per Lorentz factor in the delayed ejecta. The minimum Lorentz factor $\Gamma_m$ of 
the ejecta determines the observer time when the injection ends. 
A sudden energy input (\ie large parameter $s$), resembling the collision of a 
second shell with the leading fireball, may lead to a temporary flattening of $\Gamma$
as a function of $r$, as shown by the small value of $-\dd \log \Gamma/\dd \log r$
at $T \sim 3$ days for $s=10$. The flux of the synchrotron radiation emitted by the 
remnant at a frequency $\nu$ above the synchrotron peak $\nu_p$ (of $\nu F_{\nu}$, 
the power-per-decade) is proportional to 
$\Gamma^{8+4\beta}T^3$ if the electrons radiating at $\nu$ are adiabatic, and 
proportional to $\Gamma^{4+4\beta}T^2$, if the same electrons are radiative, where 
$\beta$ is the slope of the spectrum above $\nu_p$: $F_{\nu} \propto \nu^{-\beta}$.
For $\beta \sim 1$, as observed in most afterglows, the remnant flux varies like 
$\Gamma^{12} T^3$ and $\Gamma^8 T^2$ for adiabatic or radiative electrons, respectively. 
This means that the afterglow corresponding to the remnant evolution shown in Figure 2 
for $s=10$ should exhibit a substantial brightening, with $F_{\nu}$ increasing as 
fast as $T^3$ (adiabatic electrons) or $T^2$ (radiative electrons) at $T \sim 3$ days.

\subsection{Conical Remnants}

 Before the effect of the sideways expansion becomes important, the bulk Lorentz factor 
of an adiabatic jet-like remnant running into and a homogeneous external medium is given 
by 
\begin{equation}
 \Gamma = \Gamma_d (r/r_d)^{-3/2} \;.
\label{Gadb}
\end{equation}
The radius $r_j$ at which the remnant Lorentz factor has decreased to $\theta_0^{-1}$, 
\ie when an observer located on the jet's symmetry axis "sees" the jet's edge, if the
jet sideways expansion until $r_j$ is ignored, is given by 
\begin{equation}
 r_j=(\Gamma_d \theta_0)^{2/3} r_d \;.
\label{rjet}
\end{equation}
Using equations (\ref{Time}) and (\ref{Gadb}) it can be shown that  we have 
$r \propto T^{1/4}$ and $\Gamma \propto T^{-3/8}$. For $r > r_j$ the area visible to the 
observer no longer increases as $(\Gamma T)^2 \propto T^{5/4}$, as it did when $r < r_j$, 
but as $(r \theta_0)^2 \propto T^{1/2}$, thus the light-curve decay should steepen by a 
factor $T^{-3/4}$. 

 The remnant is still relativistic ($\Gamma > 2$) at $r_j$ if $\theta_0 \siml 30\deg$. 
If the jet is sufficiently narrow, then the sideways escape may have an important effect 
on the remnant dynamics before the onset of the non-relativistic phase. Taking the radius 
at which the jet's half-angle is twice the initial one as the definition of the radius 
$r_b$ where the sideways expansion becomes important, and using the equation for $r_b$
derived by Rhoads (1999), we obtain
\begin{equation}
 r_b = (75/4)^{1/3} r_j = \left[ ({75}/{8}) \Gamma_0^2 \theta_0^2 \right]^{1/3} r_d \;.
\label{rb}
\end{equation}
Equation (\ref{rb}) is valid only if the remnant is still relativistic at $r_b$. 
Since the remnant Lorentz factor at $r_b$ is $\Gamma_b=(2/5\sqrt{3})\theta_0^{-1}$ 
(Rhoads 1999), this condition reduces to $\theta_0 \ll 0.1\,{\rm rad} \sim 6 \deg$ 
($\Omega_0 \ll 4 \times 10^{-2}$ sr). 

 Equations (\ref{rjet}) and (\ref{rb}) show that $r_b/r_j \sim (75/4)^{1/3} \sim 2.7$, 
where we used $\Gamma_d=\Gamma_0/\sqrt{2}$, for consistency with Rhoads' results, which 
is close to the value $0.62 \Gamma_0$ derived by PMR98. Therefore the jet edge effect 
should always be seen before that of the sideways expansion. Since $T \propto r^4$ for
$r < r_b$, the ratio of the observer times at which the sideways expansion and jet edge 
phases begin should be $T_b/T_j = (r_b/r_j)^4 \sim 50$. 

 As shown by Rhoads (1999), during the sideways escape phase $\Gamma$ decreases 
exponentially with radius: 
\begin{equation}
 \Gamma = \Gamma_b e^{-(r-r_b)/r_e}\;.
\label{Gexp} 
\end{equation}
The exponential constant can be cast in the form $r_e = (\Gamma_0 \theta_0)^{2/3} r_d =
0.47\, r_b$. Thus during the exponential regime
\begin{equation}
 -\dd \ln \Gamma / \dd (r/r_d) = (\Gamma_0 \theta_0)^{-2/3} \;.
\label{exp}
\end{equation}
 
 With the aid of equations (\ref{Time}) and (\ref{Gexp}) one can show that in the
exponential regime $\Gamma \propto T^{-1/2}$, thus the non-relativistic phase 
begins at $T_{nr} = (1/4) \Gamma_b^2 T_b = (75\,\theta_0^2)^{-1} T_b$. 
Using equation (\ref{rdec}), the times $T_j$, $T_b$ and $T_{nr}$ can be calculated:
\begin{equation}
 50\,T_j = T_b = (75\,\theta_0^2) T_{nr} = 1.0\, \left(\frac{1+z}{2}\right) 
   \left(\frac{\varepsilon_{0,54}}{n_0}\right)^{1/3} \Omega_{0,-3}^{4/3}\, [{\rm day}] \;,
\label{TjTb}
\end{equation}
where $\varepsilon_{0,54}$ is the initial energy per solid angle in units of $10^{54}\, 
{\rm erg\, sr^{-1}}$, $n_0$ is the external medium number density in ${\rm cm^{-3}}$ 
and $\Omega_0 = 10^{-3}\,\Omega_{0,-3}$ sr.

 The dynamics of adiabatic conical remnants is shown in Figure 3, where we assumed a 
homogeneous external medium.  As can be seen the exponential regime (\ie the flattest 
part of each curve) is less evident for ejecta whose solid angle is larger 
than $\sim 10^{-2}$ sr, when the onset of the non-relativistic regime occurs before 
the sideways expansion has a significant effect on the remnant dynamics. For a jet
with $E_0 = 10^{51}$ ergs and $\Omega_0=10^{-3}$ sr ($\theta_0 = 1\deg$), which
the case shown in Figure 3 with dotted lines, equation (\ref{TjTb}) predicts that the 
jet edge is seen at $T_j = 0.5$ hours, the exponential regime starts at $T_b = 1.0$ day 
and ends at $T_{nr} = 44$ days. Numerically we obtain that $\Gamma = \theta^{-1}$ at 
$T_j = 1.0$ hours, $\theta$ being the jet half-angle, $T_b = 0.35$ days and $T_{nr} = 37$ 
days for {\sl Model 1}, and 1.2 hours, 0.35 days and 41 days, respectively, for 
{\sl Model 2}. Note that the numerical and analytical results are in good agreement 
for $T_{nr}$. 

 The discrepancy for the $T_j$ and $T_b$ values arises from the fact 
that in the analytical derivation the effect of the sideways expansion on the remnant 
deceleration during the power-law phase was ignored. Because there is some sideways 
expansion during this phase, the jet half-angle $\theta$ is increasingly larger than 
$\theta_0$, and $\Gamma$ drops below $\theta^{-1}$ after it has reached the value 
$\theta_0^{-1}$, thus the analytical $T_j$ underestimates the numerical one. Numerically
we found that when $\Gamma = \theta^{-1}$ the jet angle is $\theta_j = 1.2\, \theta_0$. 
In the analytical treatment presented by Rhoads (1999) the increase in the swept-up mass 
due to the jet broadening during the power-law phase is ignored, which means that, for 
the same radius, the analytical $\Gamma$ is larger than the numerical one, thus the
analytical comoving time and the jet angle at given $r$ are smaller than the values 
obtained numerically. Therefore the analytical $T_b$ overestimates the time when 
$\theta = 2\,\theta_0$. Numerically we found that at $T_b$ given by equation (\ref{TjTb}) 
the jet angle is $\theta = 2.5\,\theta_0$. 

 For the jet whose dynamics is shown in Figure 3 ($\theta_0 = 1\deg$, $\Gamma_0=500$), 
equation (\ref{exp}) predicts that during the exponential phase $-\dd \ln \Gamma/
\dd (r/r_d)=0.23$, which is less than the values shown in Figure 3 at times after
$T_b$ and before $T_{nr}$: $0.33 \pm 0.08$ for {\sl Model 1} and $0.30 \pm 0.07$ 
for {\sl Model 2}. This is consistent with the fact that  $-\dd \ln \Gamma/
\dd (r/r_d) \propto r_e^{-1} \propto r_b^{-1}$ and that the numerical $r_b$ is smaller
than the analytical one. 

 The same conclusions have been reached by comparing the analytical and numerical 
results for other values of $\Omega_0$ and $\Gamma_0$, the most important them being 
that $T_b/T_j = 7 \div 10$, which is $5 \div 7$ times smaller than obtained analytically.

\section{Synchrotron Light-Curves from Beamed Ejecta}

 The calculation of the afterglow light-curve is described in PMR98, and consists 
in integrating the remnant synchrotron emission over its dynamical evolution, over the 
electron distribution in each infinitesimal layers of swept-up external gas and over 
the angle relative to the jet axis. The electron distribution is initialized as a 
power-law and evolved through synchrotron and adiabatic cooling. We assume an electron 
index $p=3$, and that electrons acquire 10\% of the
internal energy after shock acceleration. We also assume a turbulent magnetic field
which stores $10^{-4}$ of the shocked gas internal energy, \ie the magnetic field
intensity is two orders of magnitude weaker than the equipartition value, which
leads to negligible radiative losses and an adiabatic remnant evolution. For a more
general case where there could be substantial radiative losses, some of the remnant
dynamics equations are altered as following. The factor $A$ in equation (\ref{dUex}) 
is set to 1 and a new term is added to equation (\ref{dU}), representing the radiative 
losses. The second term in the right hand side of equation (\ref{dU}), representing
the adiabatic losses, is split into a term for the adiabatic losses of the baryons 
and another one for the electrons. The electronic adiabatic losses act simultaneously
with the radiative losses and are put together in a differential equation for
electron cooling (equation [13] in PMR98).

 The calculation of the photon arrival time takes into account both the geometrical
curvature of the remnant (photons emitted by the fluid moving off the observer's
line of sight toward the center of the shell arrive later than those emitted along
this central line of sight) and the thickness of the shocked fluid, whose light
crossing-time is not negligible compared to the observer time since the main burst. 
Here we assume that there is a negligible mixing behind the forward shock, such that 
the electron distribution in any remnant infinitesimal layer is the one injected when 
that layer was added to the shocked structure, evolved through adiabatic and radiative 
losses. The shell thickness at the time when a new infinitesimal layer of swept up 
external gas is added can be obtained from the remnant comoving volume, whose calculation
is described in \S2.4. 

 The effect of the sideways expansion on the optical afterglow seen by an observer 
located on the jet axis is shown in Figure 4$a$. The initial energy per solid angle 
is the same for all remnants, only the jet initial half-angle $\theta_0$ is changed.
The afterglow brightness should be independent of $\theta_0$ until $T_j$, when the 
flow Lorentz factor has become sufficiently low that the observer sees the edge of 
the jet. This feature is better seen if the sideways expansion is "switched off",
because in the case where it is taken into account there is a non-negligible jet 
broadening until $T_j$. 

 For the afterglows shown in Figure 4$a$ $\Gamma$ decreases to
$\theta^{-1}$ at $T_j = 0.005$, 0.086, and 1.8 days for $\theta_0 = 1\deg$, $3\deg$, 
and $9\deg$, respectively, if the jet broadening is not taken into account (\ie 
$\theta=\theta_0$ at all times), and 0.009, 0.16, and 4.1 days, respectively, if the 
sideways expansion is accounted for. These are the times when photons emitted from the 
forward shock along the remnant center--observer line (which is the jet axis in this case) 
arrive at the observer. Photons emitted from the fluid located closer to the contact 
discontinuity arrive up to twice later. For $T < T_j$, photons emitted from the forward 
shock regions moving at an angle $\Gamma^{-1}$ off this central line of sight arrive 
at $T=(1-\cos \Gamma^{-1}) (r/c) \sim (2\Gamma^2)^{-1} (r/c)$, which is factor 8
larger than the arrival time from the forward shock $T_{FS}=(16\Gamma^2)^{-1} (r/c)$,
as can be shown using equations (\ref{Time}) and (\ref{Gadb}). 

 The times when the half-angle of the jets whose afterglows are shown in Figure 4$a$  
reach twice their initial values are $T_b = 0.08$, 1.3, and 27 days, for $\theta=1\deg$, 
$3\deg$, and $9\deg$, respectively. The optical light-curves shown in Figure 4$a$ steepen 
smoothly around $T_b$, while the light-curves of the non-broadening jets maintain the 
decay slopes they had before $T_b$ (of course, $T_b$ has no meaning for a jet of 
constant opening). The light-curve steepening that can be seen for the non-broadening 
$\theta_0=9\deg$ jet around $T=10$ days is due to the passage of the cooling break 
through the optical band. 

 It can be noticed that the slopes of the light-curves for non-broadening jets shown 
in Figure 4$a$ are not constant after the $T_j$'s given above and before $T_b$, as the 
remnant geometrical curvature delays the photon arrival time from regions off the jet 
axis. Moreover, the received power per solid angle being proportional to 
$[\Gamma (1-\beta \cos \delta)]^{-4}$, where $\delta$ is the angle relative to the 
central line of sight at which an infinitesimal emitting region moves, implies that at 
$T_j$ this power per solid angle from the jet edge ($\delta = \theta_j = \Gamma_j^{-1}$) 
is 16 times smaller than that from $\delta = 0$, which leads to the conclusion that it 
should take longer than just $T_j$ to see an afterglow dimming rate in excess of that 
existent until $\sim T_j$ (this is confirmed by the numerical results, as shown below). 
Put together with the fact that $T_b \sim (7\div 10) T_j$ (see previous section) this 
suggests that the effects arising from seeing the jet edge and from the sideways 
expansion may not be so clearly distinguishable for the observer.

 For the sideways expanding jets shown in Figure 4$a$, the non-relativistic phase 
begins at $T_{nr} = 11$, 18 and 32 days (forward shock times) for $\theta_0= 1\deg$, 
$3\deg$, and $9\deg$, respectively, while for the jets where sideways expansion was
not taken into account in the dynamics $T_{nr}=59$ days, independent of $\theta_0$.

 The effect of the sideways expansion in the case where the observer is located
off the jet axis, at an angle $\theta_{obs}$ relative to this axis, is illustrated 
in Figure 4$b$. The major difference from the $\theta_{obs} = 0$ case is that,
shortly after the light-curve rises, which happens later for larger $\theta_{obs}$, 
the broadening of the jet yields a brighter afterglow than in the case where the
sideways expansion is ignored. This is due to the fact that for broadening jets there
is some shocked fluid approaching the observer line of sight toward the remnant
center, along which the relativistic effects are maximal, while for non-broadening
jets it is only the decrease of the remnant bulk Lorentz factor that "brings" the
observer into the cone of the relativistically beamed radiation and thus to see the 
afterglow.

 In Figure 5 we illustrate the importance of integrating the remnant emission over 
the angle relative to the observer and by taking into account the finite thickness 
of the source, \ie the time delays introduced by the location within the remnant 
where the radiation is released.
The curves shown with thick broken lines correspond to afterglows where both of the 
above effects are taken into account. The curves shown with thin broken lines 
represent afterglows where the integration over the polar angle was "switched off",
in the sense that the light-curve has been calculated as if all the emitting fluid
were moving directly toward the observer.  However, to account for the fact that
at times $T < T_j$ and in the presence of the relativistic effects the observer 
receives radiation only from the fluid moving within $\Gamma^{-1}$ off the central
line of sight, the afterglow's brightness at $T < T_j$ has been corrected by a factor 
$(\Gamma \theta)^{-2} \sim (\Gamma \theta_0)^{-2}$. As can be seen, ignoring the
effect arising from the geometrical curvature of the source leads to a substantial
overestimation of the early radio afterglow (Figure 5$a$), introduces errors of more 
than 1 mag in the early optical afterglow and of $\sim 1/2$ mag at later times 
(Figure 5$b$), and underestimates the X-ray afterglow by a factor $\siml 4$. 

 The evolutions of the slope of the afterglows obtained by accounting for the shell 
curvature are given in Figure 5$d$, where we also show with vertical lines the
important forward shock times. For the parameters chosen here for the electron 
fractional energy and for the magnetic field intensity, the spectral cooling break 
is always between optical and X-ray and above the synchrotron peak, thus the electrons 
radiating at this peak, in radio and in optical are adiabatic, while those radiating 
in X-ray are radiative. Following the analytical treatment of \Mesz \etal (1998), 
one can derive the following asymptotic afterglow light-curves for the case discussed 
here:
\begin{equation} 
 F_R \propto \left\{ \begin{array}{ll} 
     T^{1/2} & T < T_j       \\  T^{-1/4} & T_j < T < T_b     \\ 
     T^{-1/3} & T_b < T < T_R  \\  T^{-p} & T_R < T < T_{nr}
   \end{array} \right.
\label{FR}
\end{equation}
\begin{equation} 
 F_O \propto \left\{ \begin{array}{ll} 
    T^{1/2} & T < T_O        \\ T^{-3(p-1)/4} & T_O < T < T_j    \\ 
    T^{-3p/4} & T_j < T < T_b  \\ T^{-p} & T_b < T < T_{nr}
   \end{array} \right.
\label{FO}
\end{equation}
\begin{equation} 
 F_X \propto \left\{ \begin{array}{ll}  T^{-(3p-2)/4} & T <T_j    \\  
     T^{-(3p+1)/4} & T_j < T < T_b \\ T^{-p} & T_b < T < T_{nr}
    \end{array} \right.
\label{FX}
\end{equation}
where $p=3$ for the light-curves shown in Figure 5. The afterglow slope is expected 
to steepen at all frequencies by $-3/4$ when the jet edge is seen, while the changes 
in the same slope due to the jet sideways expansion are expected to be $-1/12$ in 
radio, $-p/4=-3/4$ in optical, and $-(p-1)/4=-1/2$ in X-ray. It should be noted here
that in the derivation of equations (\ref{FR})--(\ref{FX}) the time evolution of the
remnant specific intensity at the synchrotron peak was calculated assuming that all
the electrons radiate at this peak (the electrons being adiabatic), which is a rough 
approximation given that the injected electron distribution (and implicitly the peak
energy of their synchrotron emission) changes substantially during an adiabatic 
timescale, which is comparable to the age of the remnant.

 The afterglow slopes shown in Figure 5$d$ for the case where the "angular effect" is
ignored exhibit the trends predicted by equations (\ref{FR})--(\ref{FX}). The largest
difference is shown by the optical light-curve at times between $\sim T_j$ and until 
after $T_b$. It results mostly from the fact that at these times the optical band is 
above but close to the synchrotron peak, the integration over the electron distribution 
in the shell leading to a spectrum that does not exhibit a pure $-(p-1)/2$ slope (the 
electrons radiating in optical are adiabatic), as assumed in the derivation of the 
analytical light-curves, but to a smooth transition between the $1/3$ slope below the 
synchrotron peak and the $-p/2$ slope above the cooling break (see Figure 6).

 Note from Figure 5$d$ that taking into account the curvature of the emitting shell 
delays the jet edge break, as expected, until times comparable to $T_b$. The effect of
the remnant thickness on the afterglow slope evolution (not shown in Figure 5$d$) is 
similar, but weaker, to that of the remnant curvature.  Figure 6 shows that the spectrum 
calculated by taking into account the time delays and Lorentz boosting factors over a 
curved shell is harder than if the shell is assumed planar. A photon emitted by the gas 
moving off the observer's central line of sight is at most twice less blueshifted than 
one emitted along this line, nevertheless the former was radiated at a smaller radius, 
when $\Gamma$ and the synchrotron peak frequency (which scales as $\Gamma^4$) where 
larger, the end result of the integration over the equal arrival time surface being a 
harder spectrum. 

 The upper three panels of Figure 5 also show the light-curve obtained with {\sl Model 1}
if it is assumed that all the radiation is emitted from the location of the forward
shock, \ie if the light travel time across the source is neglected. The differences
are minor in the X-ray because the electrons emitting at high energies are radiative
and cool fast, thus the X-ray emission from the fluid that is not located close to 
the forward shock is negligible, but significant differences can be seen in the radio and
optical light-curves, as the electron radiating in these bands are adiabatic, and thus
occupy the entire volume of the remnant.
Also shown in Figures 5$a$ and 5$b$ with continuous lines are the afterglows calculated 
with {\sl Model 2} for the remnant comoving volume. Note that the largest difference 
between the optical light-curves within two models amounts to $\sim$ 1 mag in the late
afterglow. These differences are due mostly to the prescription for the comoving volume, 
which leads to different magnetic fields and shell thicknesses, and not to the adiabatic 
losses.

\section{Discussion}

 We have presented an analytical treatment for the dynamics of an expanding fireball, 
capable of following its evolution from the onset of the deceleration phase ($r \sim r_d$) 
until arbitrarily large times. The differential equations for the remnant dynamics 
given here are valid in any relativistic regime. The major assumption underlying the 
analytical derivations is that, at any time, the remnant is axially symmetric.

 This analytical treatment takes into account a possible delayed energy input 
resulting from an impulsive but uneven deposition of energy in the ejected material. 
For beamed ejecta, it also takes into account the intensification of the remnant 
deceleration due to the increase of the solid angle of the remnant and, thus, of 
the rate at which it sweeps up external gas. The results presented in the previous 
section illustrate the effect of these two factors. The treatment of the adiabatic 
losses and the calculations of the magnetic field and remnant thickness require a
prescription for how to calculate the remnant volume.
We considered two models for this: {\sl Model 1} is based on the assumption that, 
if the accumulation of swept-up gas is subtracted, the remaining increase of the lab 
frame volume is due only to the $r^2$ increase of the remnant area, and {\sl Model 2}, 
which is based on the assumption that the density profile behind each shock is uniform.
The two models for the comoving volume calculation lead to significant differences in 
remnant dynamics when there is a sharp delayed energy input, as shown in Figure 2, and
in the case of beamed ejecta, as shown in Figure 3. There are also significant 
differences among the radio and optical afterglows calculated within the two models
for the remnant volume, as illustrated in Figures 5$a$ and 5$b$.

 We have compared the features of the dynamics of beamed ejecta calculated numerically
with some analytical results derived from those presented by Rhoads (1999).  The most 
important difference is that the ratio of the observer times $T_b$, when the jet 
half-angle $\theta$ reaches twice its initial value due to the sideways expansion, and 
$T_j$, when the remnant Lorentz factor has decreased to $\theta^{-1}$, is analytically 
overestimated to $\sim 50$, while the numerics lead to a value $\siml 10$. $T_j$ is the 
time when the afterglow is expected to steepen by $T^{-3/4}$ due to the finite angular 
opening of the ejecta, while at $T_b$ the afterglow should further steepen by at most 
$T^{-p/4}$, where $p$ is index of the injected electron distribution, due to the 
intensification of the remnant deceleration caused by the continuous jet broadening. 
The afterglow slope evolutions shown in Figure 5$d$ suggest that, due to the remnant 
curvature, the $T_j$ defined above actually underestimates the time when the finite 
angular extent of the remnant yields a substantial steepening of the light-curve, and 
that jet edge break overlaps with the weaker light-curve steepening due to the jet 
sideways expansion.

 We have compared the numerical light-curves with the asymptotic behaviors expected 
analytically and we have analyzed the importance of integrating the remnant emission 
over its angular opening, as the Doppler boosting and the photon arrival time are 
functions of the angle between the direction of motion of the emitting fluid region 
and the direction toward the observer. Figure 5 shows that this effect is quite 
important, and should be taken into account for accurate afterglow calculations.
We have also found that the thickness of the source is an important factor, though
less than the shell curvature, in the calculation of radio and optical afterglows.
Lastly, the afterglow spectrum at frequencies around the peak and the cooling breaks,
and between these breaks (if they are not too far from each other), may be poorly
approximated as a power-law, particularly if the electrons radiating at the frequency 
of interest are adiabatic (see Figure 6), implying that they have been injected over 
a wide range of radii and bulk Lorentz factors. In this case, a numerical integration
of the electron injection and cooling is necessary to obtain an accurate afterglow
emission.

\acknowledgements{This research has been partially supported through NASA
                  NAG5-2857 and NAG5-3801.}

\onecolumn
\input PSfig

\begin{figure}
\centerline{\psfig{figure=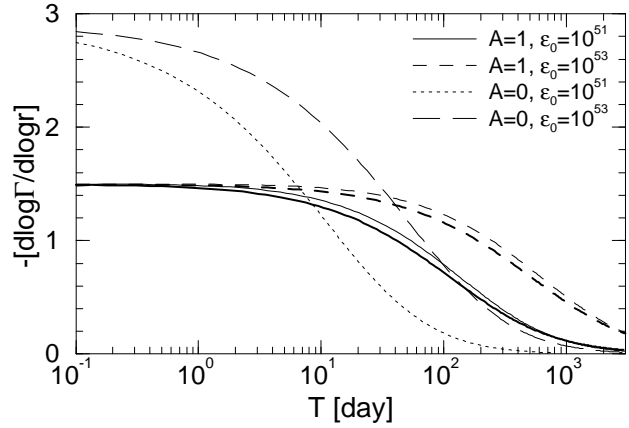}}
\vspace*{-8cm}
\caption{Evolution of $-\dd \log \Gamma/\dd \log r$ for adiabatic ($A=1$) and 
   radiative ($A=0$) spherical remnants, with no delayed injection and homogeneous 
   external medium. Parameters: $z=1$, $\Gamma_0=500$, $n_d=1\,{\rm cm^{-3}}$ and 
   $\varepsilon_0 \equiv E_0/\Omega_0$ as given in the legend, in units of 
   ${\rm erg\, sr^{-1}}$. Thick curves correspond to {\sl Model 1}, while thin 
   lines are for {\sl Model 2}. Obviously, the numerical results are the same 
   for both models of adiabatic losses if the remnant is fully radiative ($A=1$).
   Note that a larger energy per solid angle in the ejecta leads to a longer 
   relativistic phase.} 
\end{figure}

\begin{figure}
\vspace*{-1cm}
\centerline{\psfig{figure=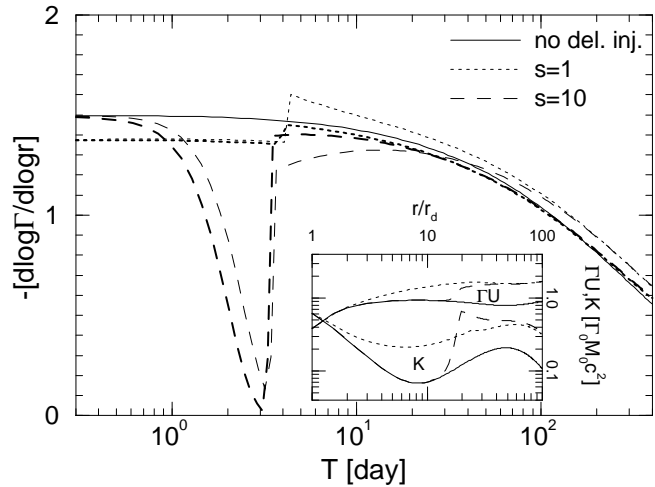}}
\vspace*{-7cm}
\caption{Effect of a power-law delayed energy input on the dynamics of an 
   adiabatic spherical remnant interacting with a homogeneous external medium. 
   The inset shows the evolution of the kinetic and lab frame internal energy, in 
   units of initial remnant energy. The parameters of the injection are $\Gamma_m=10$, 
   $E_{INJ}=E_0$ and $s$ is given in the legend. Other parameters are: $z=1$, 
   $\varepsilon_0 = 10^{52}\, {\rm erg\, sr^{-1}}$, $n_d=1\, {\rm cm^{-3}}$. 
   As for Figure 1, {\sl Model 1} solutions are shown with thick lines, while 
   {\sl Model 2} solutions are indicated with thin curves.  The continuous thin 
   curve is for the case with no delayed injection (added for comparison). 
   Note that the differences between the solutions obtained with the two models 
   for adiabatic losses are larger at times when most of the injection 
   takes place, indicating that the adiabatic cooling of the delayed shocked ejecta 
   is the source of these differences.}
\end{figure}

\begin{figure}
\vspace*{-5cm}
\centerline{\psfig{figure=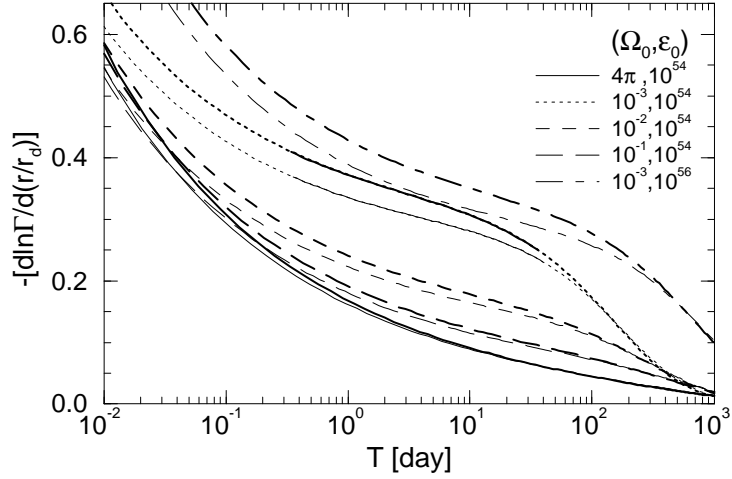}}
\vspace*{-12cm}
\caption{Evolution of $-\dd \ln \Gamma/\dd r$ for adiabatic beamed ejecta. 
   The remnant dynamics depends on the initial solid angle of the ejecta (given in
   the legend in steradians) and on the initial energy per solid angle (given in the 
   legend in ${\rm erg\, sr^{-1}}$). Other parameters: $z=1$, $n_d=1\, {\rm cm^{-3}}$, 
   $\alpha=0$, $\Gamma_0=500$. The meaning of thick and thin curves is the same as 
   for Figures 1 and 2. The parts indicated with a continuous curve for the ($\Omega_0=
   10^{-3}$, $\varepsilon_0 = 10^{54}$) models correspond to $T$ between $T_b$
   and $T_{nr}$, \ie when the remnant opening angle is more than twice the initial 
   one and the remnant is still relativistic ($\Gamma > 2$). }
\end{figure}

\begin{figure}
\vspace*{-13cm}
\centerline{\psfig{figure=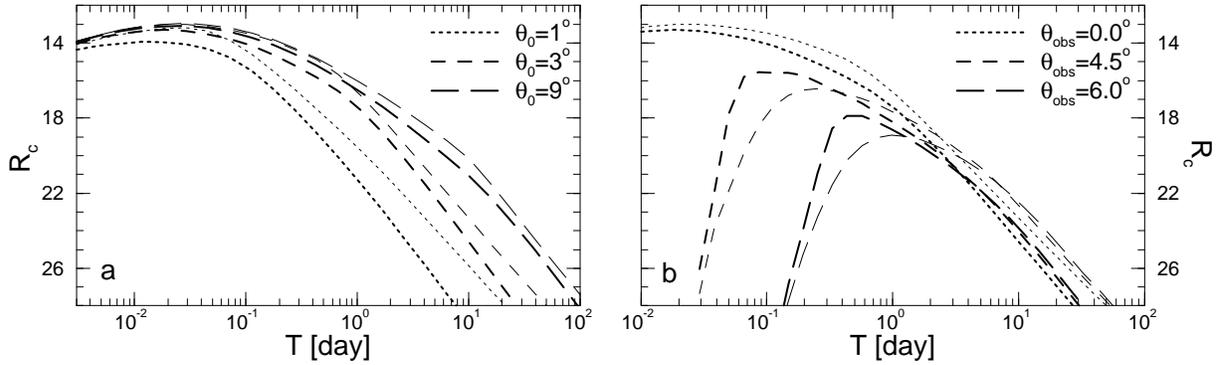}}
\vspace*{-12cm}
\caption{Optical light-curves and the effect of sideways expansion for beamed ejecta.
   For both panels the $R_c$ magnitudes are calculated with {\sl Model 1}.  
   Thick curves are for the light-curves obtained by taking into account the sideways
   expansion of the ejecta, both in the dynamics and the integration of light, while
   thin curves are for the case where the jet broadening is ignored.
   Parameters: $z=1$, $n_d=1\,{\rm cm^{-3}}$, $\alpha=0$ and $\varepsilon_0 = 10^{53}\, 
   {\rm erg\, sr^{-1}}$. The electron and magnetic field energy densities are 
   $10^{-1}$ and $10^{-4}$ of the internal energy density. The distribution of the 
   shock-accelerated electrons is assumed to be a power-law of index $-3$.
   Panel $a$: the observer is located on the axis of the jet, whose initial half-angle 
    is given in the legend. Note that the jet progressive broadening leads to dimmer
    afterglows. 
   Panel $b$: a jet of initial half-angle $\theta_0 = 3\deg$ is seen at different 
    angles. Note that for $\theta_{obs} > \theta_0$ the sideways expansion yields 
    afterglows that have brighter peaks than in the case where the jet broadening 
    is not taken into account.}
\end{figure}

\begin{figure}                                                                
\centerline{\psfig{figure=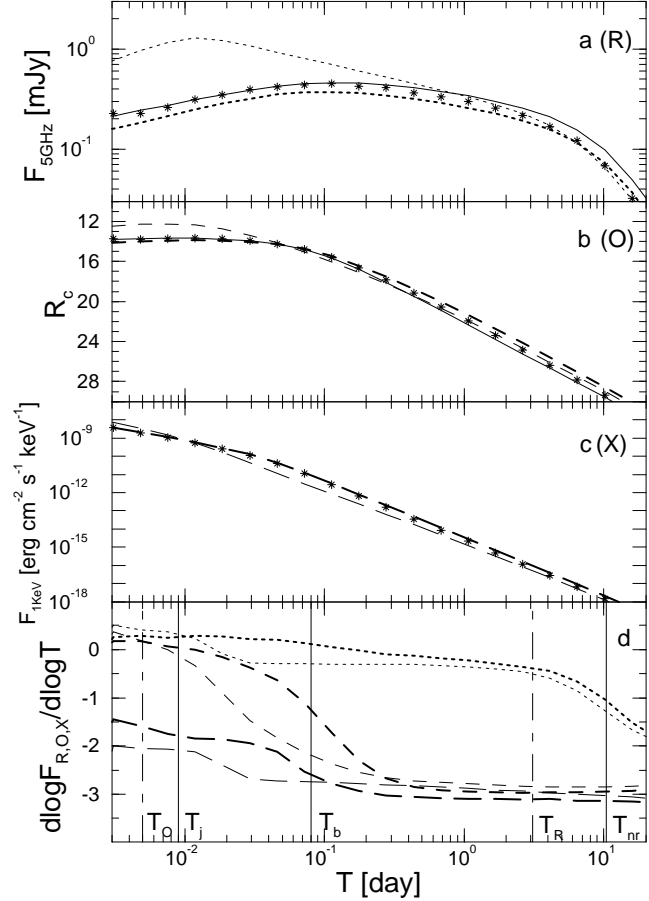}}
\vspace*{2cm}
\caption{Effects of the remnant geometrical curvature and thickness on the observed 
   light-curve. The upper three panels show the {\sl Model 1} radio, optical and X-ray 
   afterglows for a jet with $\theta_0 = 1\deg$ seen by an observer located on its 
   symmetry axis. Other parameters are as for Figure 4. The light-curves whose
   calculations took into account both the source curvature and thickness are shown 
   with thick broken lines, those for which the curvature was ignored are represented 
   with thin broken lines, while stars are for the case where it was assumed that all 
   the radiation is emitted from the location of the forward shock.
   The evolution of the radio (dotted line), optical (dashes), and X-ray (long dashes)
   light-curve slopes for the first two cases are shown in panel $d$. The relevant 
   forward shock times are indicated on the abscissa. $T_O = 0.005$ days and $T_R = 3.1$ 
   days are the times when the synchrotron peak passes through the optical and radio 
   bands, respectively. For comparison, panels $a$ and $b$ also show with continuous 
   thin lines the light-curves obtained with {\sl Model 2}. The X-ray afterglows in the 
   two models for the comoving volume are indistinguishable.}
\end{figure}                                                             

\begin{figure}                                                                
\centerline{\psfig{figure=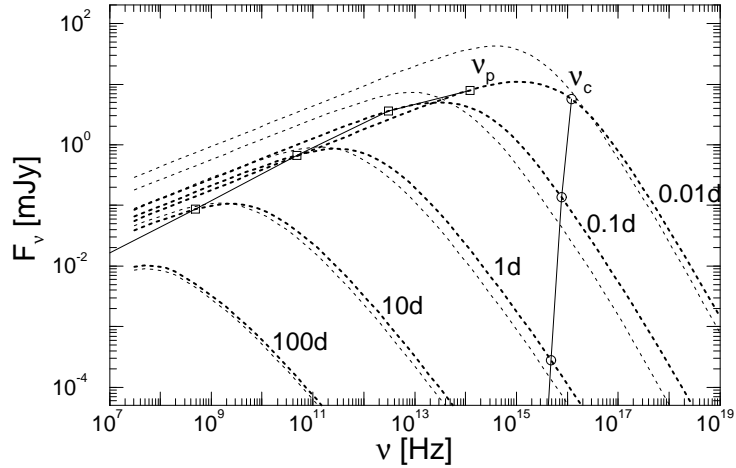}}
\vspace*{2cm}
\caption{Spectral evolution of the afterglow shown in Figure 5 and the effects of the 
   remnant curvature on the observed spectrum. Thick dotted curves are for spectra whose
   calculation included the shell curvature, while thin dotted lines are for the case
   where it was ignored. The squares connected by a continuous line indicate the synchrotron
   peak $\nu_p$ of the radiation integrated over the power-law distribution of the electrons 
   freshly injected by the forward shock, while circles show the peak $\nu_c$ of the radiation 
  from the electrons that cool radiatively on a timescale 
   equal to that of the adiabatic losses. In both cases the Doppler factor corresponding to 
   the motion toward the observer (the jet axis in this case) was used. From right to left 
   the observer times for each spectrum are: 0.01, 0.1, 1, 10, and 100 days. Note that 
   the integration over the equal arrival time surface yields harder spectra and spectral 
   peaks above $\nu_p$. Also note the smooth transition from 
   the $1/3$ power-law spectrum at $\nu < \nu_p$ to the $-p/2=-3/2$ one at $\nu > \nu_c$, 
   and the absence of a perfect $-(p-1)/2=-1$ power-law for $\nu_p < \nu < \nu_c$.}
\end{figure}

\end{document}